\documentclass[journal]{IEEEtran}

\usepackage{graphicx}
\PassOptionsToPackage{hyphens}{url}
\usepackage{hyperref}
\usepackage{subfigure}

\begin{document}
%
\title{Blockchain-empowered Edge Intelligence for Internet of Medical Things Against COVID-19}
%
%
%

\author{Hong-Ning Dai,~\IEEEmembership{Senior Member,~IEEE,}
        Yulei Wu,~\IEEEmembership{Senior Member,~IEEE,}
        Hao Wang,~\IEEEmembership{Senior Member,~IEEE,}
        Muhammad Imran,~\IEEEmembership{Senior Member,~IEEE,}
        Noman Haider,~\IEEEmembership{Member,~IEEE}
\thanks{H.-N. Dai is with Faculty of Information Technology, Macau University of Science and Technology, Macau. email: hndai@ieee.org.}
\thanks{Y. Wu is with the College of Engineering, Mathematics and Physical Sciences, University of Exeter, Exeter, EX4 4QF, U.K. e-mail: y.l.wu@exeter.ac.uk.}
\thanks{H. Wang is with Department of Computer Science, Norwegian University of Science and Technology, Gj\o vik, Norway. email: hawa@ntnu.no.}
\thanks{M. Imran is with College of Applied Computer Science, King Saud University, Riyadh, Saudi Arabia. email: dr.m.imran@ieee.org.}
\thanks{N. Haider is with College of Engineering and Science at Victoria University, Australia. email: noman90@ieee.org.}
}

%
%

\markboth{Journal of \LaTeX\ Class Files,~Vol.~14, No.~8, August~2015}%
{Shell \MakeLowercase{\textit{et al.}}: Bare Demo of IEEEtran.cls for IEEE Journals}
%



\maketitle

\begin{abstract}
We have witnessed an unprecedented public health crisis caused by the new coronavirus disease (COVID-19), which has severely affected medical institutions, our common lives, and social-economic activities. This crisis also reveals the brittleness of existing medical services, such as over-centralization of medical resources, the hysteresis of medical services digitalization, and weak security and privacy protection of medical data. The integration of the Internet of Medical Things (IoMT) and blockchain is expected to be a panacea to COVID-19 attributed to the ubiquitous presence and the perception of IoMT as well as the enhanced security and immutability of the blockchain. However, the synergy of IoMT and blockchain is also faced with challenges in privacy, latency, and context-absence. The emerging edge intelligence technologies bring opportunities to tackle these issues. In this article, we present a blockchain-empowered edge intelligence for IoMT in addressing the COVID-19 crisis. We first review IoMT, edge intelligence, and blockchain in addressing the COVID-19 pandemic. We then present an architecture of blockchain-empowered edge intelligence for IoMT after discussing the opportunities of integrating blockchain and edge intelligence. We next offer solutions to COVID-19 brought by blockchain-empowered edge intelligence from 1) monitoring and tracing COVID-19 pandemic origin, 2) traceable supply chain of injectable medicines and COVID-19 vaccines, and 3) telemedicine and remote healthcare services. Moreover, we also discuss the challenges and open issues in blockchain-empowered edge intelligence. 

\end{abstract}


%
\IEEEpeerreviewmaketitle

\section{Introduction} 

\IEEEPARstart{T}{he} epidemic of the newly-found coronavirus disease (COVID-19) has adversely affected our common lives, overloaded medical institutions, and paralyzed aviation and tourist industries. By October 2020, there are more than 41.6 million confirmed COVID-19 cases and more than 1,100,000 deaths over 218 countries, areas, or territories across the world according to the report of the world health organization (WHO). The outbreak of the epidemic has severely overloaded existing medical institutions and healthcare facilities. As a consequence, hospitals have been overcrowded and healthcare workers have been overloaded. Meanwhile, many elderly people and patients with chronic diseases cannot be provided with immediate and proper healthcare services.  

The recent technological advances in the Internet of Medical Things (IoMT) may help to address the challenges posed by COVID-19, especially in the digitalization of medical services. IoMT connects diverse medical devices, wearable medical sensors, biosensors, and medical instruments with the Internet so as to offer a ubiquitous monitoring service on patients \cite{Yang2020.05.19.20107326}. Assisted by artificial intelligence (AI) and big data analytics technologies, massive IoMT data can be used to diagnose, recognize, analyze, and make decisions for medical practitioners and healthcare experts. 

However, both privacy leakage concerns and security vulnerabilities of IoMT systems have shrunk back public desires of the wide adoption of IoMT. On the one hand, IoMT devices typically have limited computing capabilities and storage capacity. Consequently, a common practice is to outsource IoMT data to remote cloud services, which are nevertheless possessed by services providers who may unintentionally (by mistakes) or intentionally breach the privacy of medical data. Another adverse effect of outsourcing IoMT data to remote clouds is the high latency due to data transmissions across diverse networks. On the other hand, IoMT systems are also in peril of other mischievous attacks due to the deficiency of security schemes imposed in IoMT systems.

The emerging blockchain and \emph{edge intelligence} technologies can potentially address the above challenges in IoMT. Blockchain, characterized by decentralization, immutability, and security can enhance the security of IoMT systems and protect the privacy of IoMT data. Meanwhile, edge computing, as a new computing paradigm can offload the computing tasks from remote clouds in approximation to IoMT devices so as to reduce the latency and improve privacy preservation since IoMT data can be processed and stored close to the IoMT data owner. The convergence of edge computing and AI algorithms can bestow edge computing nodes with context-aware \emph{edge intelligence}. The in-depth blockchain with edge intelligence can offer trusted edge intelligence in order to foster the wide adoption of IoMT. 

\begin{table*}[t]
\centering
\caption{Overview of IoMT for COVID-19 pandemic}
\label{tab:IoMT}
\renewcommand{\arraystretch}{1.75}
\resizebox{\linewidth}{!}{
\begin{tabular}{|c|c|c|}
\hline
\textbf{ } & \textbf{Crowd monitoring \& prevention} & \textbf{Diagnosis \& treatment} \\
\hline\hline
\textbf{Wide area} & `Pandemic' drones  & Disinfection drones, medical/delivery drones    \\
\hline
\textbf{Local area} & Health monitoring robots, indoor IoT air quality monitor & Disinfection (remote controlled or autonomous) robots, telehealth systems   \\
\hline
\textbf{Body area} & Tracing apps, wearables, e.g., wristbands or tags, for social distancing &  Social robots, wearable sensors, e.g., temperature and respiratory signs \\
\hline
\end{tabular}
}
\end{table*}

Although there are some recent studies on exploring blockchain for IoMT or machine learning/deep learning for IoMT~\cite{dai2020blockchainenabled,zhou2020deep}, most of them only consider one technology alone instead of multiple technologies together. To bridge this gap, we investigate the blockchain-empowered edge intelligence for IoMT to address the COVID-19 crisis in this article. We first briefly introduce IoMT technologies and review edge intelligence as well as blockchain technologies to explore the inclusive potentials of blockchain-empowered edge intelligence. We then present a framework of blockchain-empowered edge intelligence for IoMT, exhibiting in three different levels: trusted device intelligence, trusted edge intelligence, and trusted cloud intelligence. We next offer solutions of blockchain-empowered edge intelligence to IoMT against COVID-19 from three different perspectives: 1) monitoring and tracing COVID-19 pandemic origin, 2) traceable supply chain of injectable medicines and COVID-19 vaccines, and 3) telemedicine and remote healthcare services. Finally, research challenges and open issues are also discussed.

\section{Internet of Medical Things} 
One of the key enablers for the so-called ``Healthcare 4.0''~\cite{gatouillat2018internet} is IoMT, which connects all types of (medical or general) devices in different scales of health and public networks to facilitate health-related functionalities and processes. Important to note that we have expanded the notion of IoMT commonly found in existing literature, e.g., ``Internet of Medical Things (IoMT) designates the interconnection of communication-enabled medical-grade devices and their integration to wider-scale health networks in order to improve patients’ health''~\cite{7516690}. This is because we observe that general IoT and public networks are also participants in health-related processes, especially in this pandemic era.  

The medical context casts strict requirements on IoMT in terms of reliability, safety, and security. This becomes especially challenging due to the fast expanding nature of the COVID-19 epidemic where ethical dilemmas often arise with different alternatives to contain and isolate the spreading of the virus. In this kind of scenarios, effective, reliable and trustworthy solutions for acquiring and storing all necessary data and careful management of data access become of paramount importance.

As summarized in Table~\ref{tab:IoMT}, we look at IoMT for COVID-19 in two axes. The first axis is healthcare phases, we differentiate: a) crowd monitoring; b) diagnosis and treatment. The second axis is the scopes and areas, we differentiate: a) \textit{wide area} mainly refers to public places, especially crowded locations such as traffic hubs, sports stadiums, tourist attractions, shopping centers, and school public spaces. b) \textit{local area} refers to localized spaces such as care homes, community health centers, hospitals, homes, classrooms, and offices. c) \textit{body area}, refers to the body of a person, with or without COVID-19 symptoms. 

Contact tracing and social distancing have been essential to contain the spreading of the virus. Based on the high percentage of smart phone ownership, especially in developed areas in the world, many countries developed and announced (some enforce the use of) smartphone apps. Many of these apps make use of the global positioning system (GPS), Wi-Fi, Bluetooth modules in the smartphones to remind/enforce social distancing and tracking of person-to-person and person-to-place histories.  Wristbands and tags attached to helmets or clothing have been developed to ensure social distancing. 

Drones, robots, and surveillance devices for human officials (e.g., helmets and glasses) have been developed to both wide and local areas to monitor crowd for social distancing, temperature and respiratory signs. When infected cases are confirmed, all logged visited places should be traced and disinfected. Remote controlled or autonomous robots or drones have also been developed for these types of tasks. Data on crowd activities can then be acquired and analyzed~\cite{zhou2020deep}. As the coronavirus is airborne, air quality monitor sensors and actuators are relevant for both wide and local areas.

Telehealth systems for remote treatment of COVID-19 patients in both care homes and residential homes can relieve the pressure on hospital facilities. Social robots can help with mental issues due to isolation for COVID-19 patients.

\section{Edge Intelligence and Blockchain} 
\subsection{Edge intelligence}
\begin{figure}[t]
\centering
\includegraphics[scale=0.4]{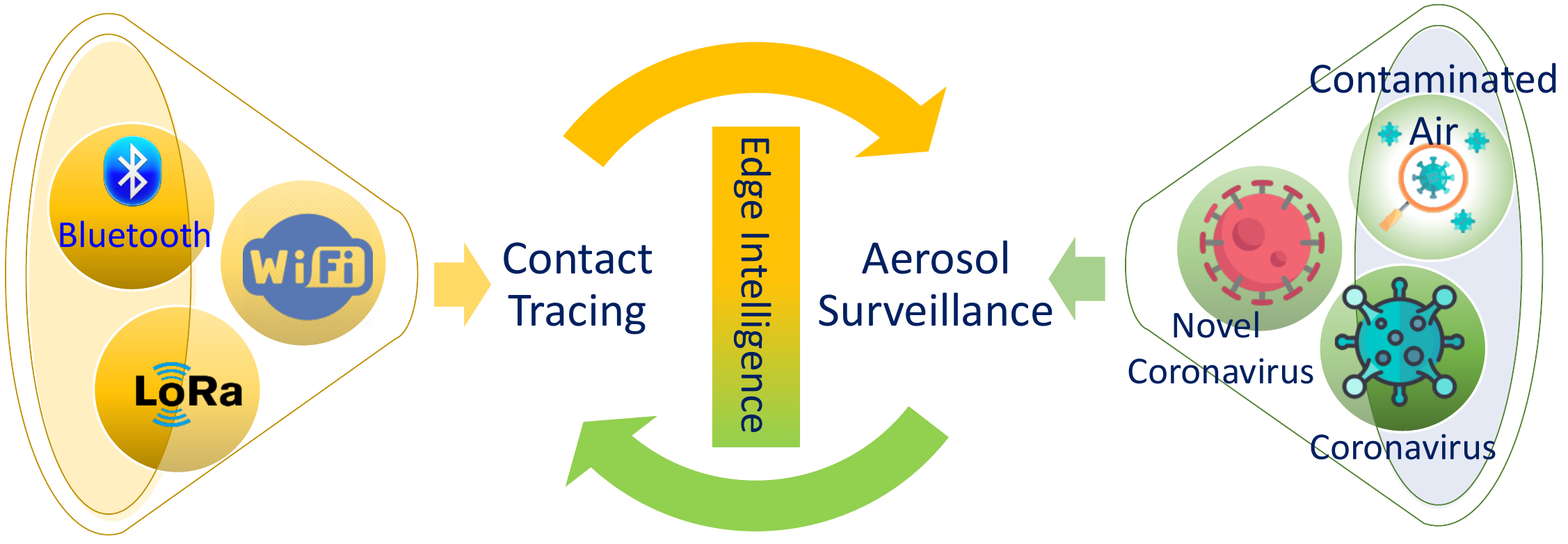}
\caption{Edge intelligence for contact tracing and aerosol surveillance.}
\label{fig:edge_intelligence}
\end{figure}

Edge computing is a computing paradigm that brings the computing capability in close proximity to where a computing task is needed. 
Edge computing, along with an increasing penetration of AI into network edge, brings the edge intelligence into diverse vertical services.
In the event of IoMT against COVID-19, edge intelligence can primarily help with the computing needs in the following two aspects: \textit{contact tracing} and \textit{aerosol surveillance}, as shown in Fig. \ref{fig:edge_intelligence}. In practice, edge computing can be deployed at IoMT gateway, Wi-Fi Access Point (AP), and evolved NodeB (eNB).

\subsubsection{Contact tracing} Bluetooth has been widely discussed and shown as one of the feasible wireless technologies for digital contact tracing. Apple\footnote{\url{https://developer.apple.com/bluetooth/}} and Google\footnote{\url{https://developer.android.com/guide/topics/connectivity/bluetooth-le}} have launched a solution with developer toolkits for iOS and Android, that include application programming interfaces (APIs) and system-level technologies to assist in enabling contact tracing. The working principle behind tracing using Bluetooth is to collect and store the Bluetooth signals broadcasting by other devices in the vicinity. The contact Bluetooth signals are then uploaded and stored in the remote cloud through an application in the IoMT devices. With a confirmed COVID-19 case, the contact IoMT devices will be identified through the stored Bluetooth signals and be notified. 

Although the Bluetooth contact tracing can in principle work, the accuracy of identifying contact IoMT devices in terms of distance between devices is a big issue. That is because Bluetooth is a peer-to-peer technology, triangulation cannot be used to achieve accurate measurement of distance. This in turn affects the performance of Bluetooth contact tracing. There are existing studies using AI techniques to improve the accuracy of distance measurement based on Bluetooth Low Energy (BLE) beacons. For example, a COVID-SAFE tool \cite{9220167}, with the cooperation of the Internet of Things (IoT), a smart phone application, and the edge intelligence for data analytics, has been developed for healthcare and physical distance monitoring.

In addition to Bluetooth, many other wireless technologies including Wi-Fi, GPS, LoRa and cellular techniques such as 4G/LTE/5G, have been used for localization and positioning in indoor and/or outdoor environment. They have been experiencing the accuracy issues due to many factors, such as low-quality of signal data. Several research works \cite{9295332} have been carried out by employing machine learning techniques, such as generative adversarial networks, to enhance the quality of signal data and thus in turn improve the accuracy of localization and positioning. Edge computing has provided a powerful nearby computing capability to reduce the computation time of machine learning algorithms and improve the data privacy issues to some extent.

\subsubsection{Aerosol surveillance} 
An aerosol is essentially a droplet in a liquid or solid form in air. A contaminated environment is a key role in transmitting the disease like COVID-19 via aerosol transmission. Aerosol transmission has been agreed as one of the main COVID-19 transmission methods and needs to be effectively monitored. On the one hand, if aerosol surveillance and analysis could be performed in a real-time manner, the transmission can be quickly intervened and limited, and the relevant people in that region can be swiftly notified. The fast analysis of aerosol transmission largely depends on the edge intelligence, which provides the necessary nearby and distributed computing capability along with the required data analytics services. On the other hand, the contact tracing solutions have not considered the possibility of a contaminated environment. Edge intelligence can effectively bridge contact tracing and aerosol surveillance and provide a more effective solution for COVID-19 pandemic mitigation.

\subsection{Blockchain}
\begin{figure}[t]
\centering
\includegraphics[scale=0.35]{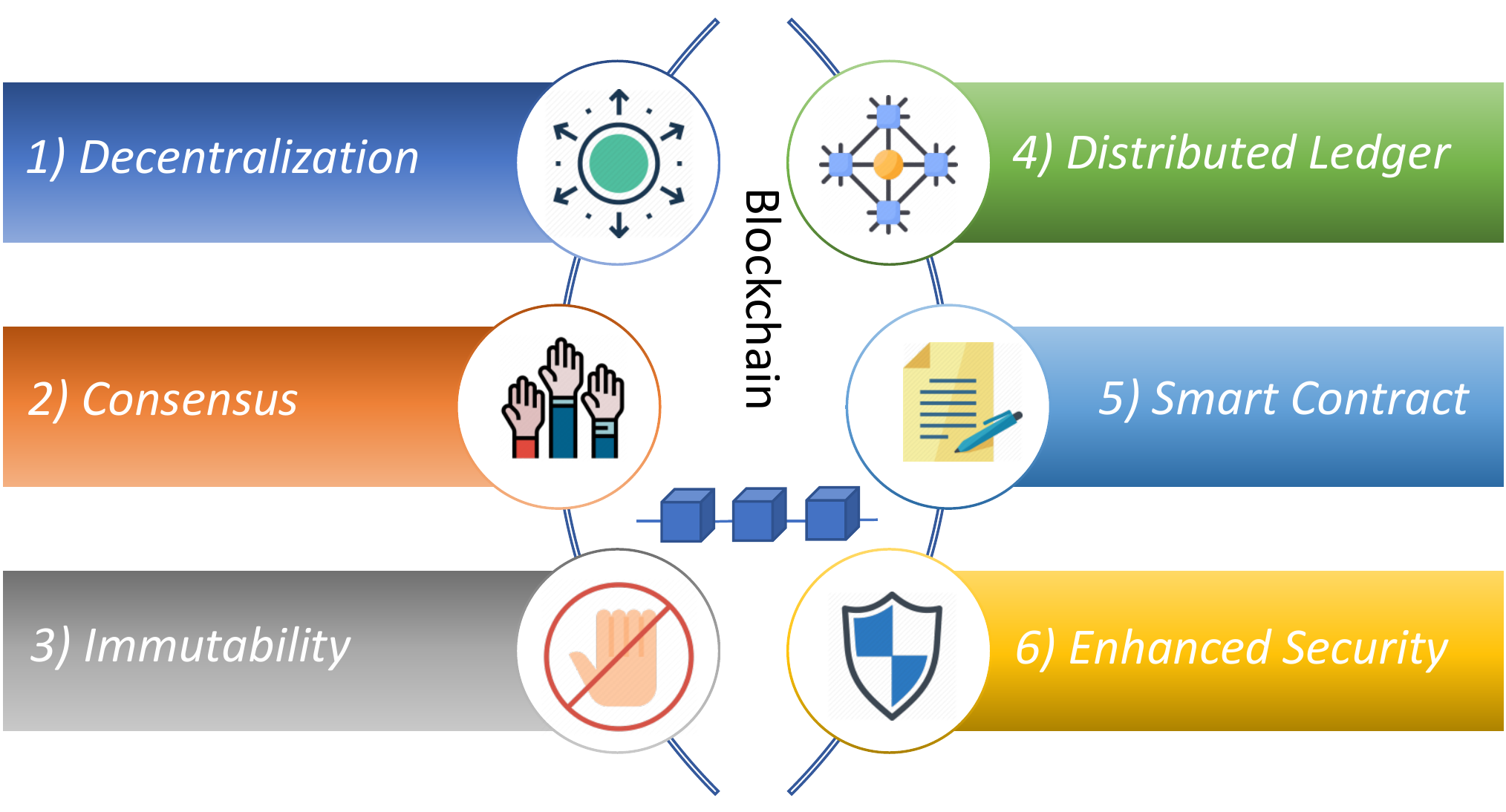}
\caption{Key features of blockchain.}
\label{fig:blockchain}
\end{figure}

Privacy is one of the biggest concerns of digital contact tracing. It is not transparent of how the collected signal data are anonymized, how they are used to track people outside the bounds of limiting COVID-19 pandemic transmission, and whether only the necessary signal data are collected. The blockchain technology has shown its merits on mitigating the privacy and many other issues in digital contact tracing. A blockchain is a peer-to-peer technology. It is essentially a distributed and secure ledger that records transactions into a chain of blocks \cite{8977439}. In what follows, how blockchain can fight against the issues encountered by digital contact tracing will be discussed. Fig. \ref{fig:blockchain} shows a summary of key features of blockchain.

\subsubsection{Decentralization} Blockchain provides decentralized computing where the participating nodes have an equal opportunity in the decision making process. It can work closely with edge computing and many peer-to-peer wireless technologies such as Bluetooth \cite{9238416} when analyzing contact tracing results.

\begin{figure*}[t]
\centering
\includegraphics[width=16cm]{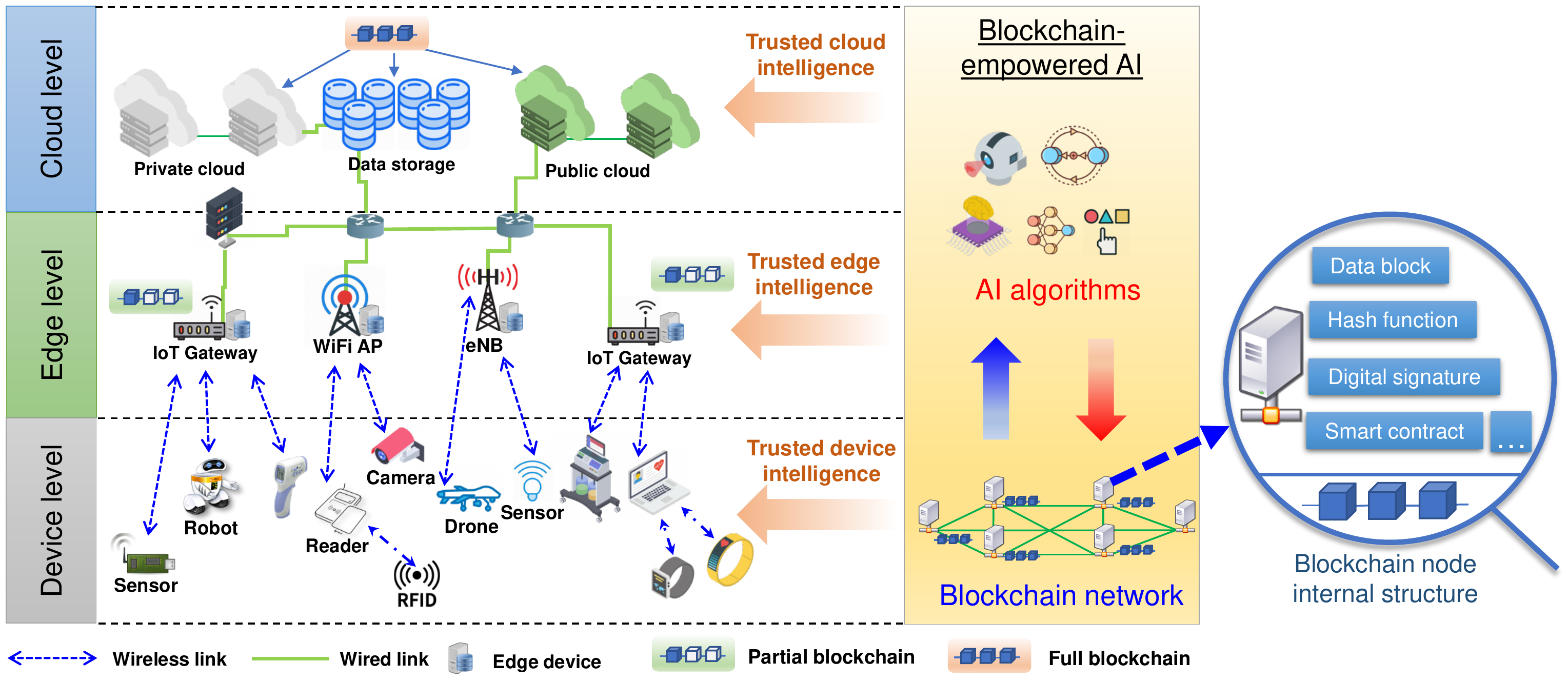}
\caption{Blockchain-empowered edge intelligence for Internet of Medical Things.}
\label{fig:blockchainEI}
\end{figure*}

\subsubsection{Consensus} Blockchain uses consensus mechanisms to achieve necessary agreement between participating nodes on the newly generated blocks such as recording contact tracing events. Consensus mechanisms can be treated as a fault-tolerant solution where mistrusting participating nodes are allowed to agree on the global state of the chain. It usually requires high computation, where edge computing is a good candidate to compensate.

\subsubsection{Immutability} All the transactions are validated and stored immutably in the blockchain, and thus the falsification to any transactions can be easily detected. This can ensure the tamper-proofing of the transactions. For example, a person's travel data that are used for contact tracing analysis is immutable.

\subsubsection{Distributed ledger} Blockchain allows a copy of the ledger of transaction records stored in each participating node. This ensures the transparency of the transactions, i.e., the use of signal data for contact tracing is transparent.

\subsubsection{Smart contract} A smart contract is a piece of software code, where it is executed automatically if electronic terms and conditions are met. The execution of each contract statement will be recorded as an immutable transaction in the blockchain. This can enable automation in many decision making in the contact tracing and notification in the event of a confirmed COVID-19 case.

\subsubsection{Enhanced security} Blockchain uses a set of encryption techniques such as PKI, digital signatures, hashing, etc. This can also provide enhanced security to the stored signal data that are used for contact tracing \cite{dai2020blockchainenabled}.

\section{Blockchain-empowered Edge Intelligence for Internet of Medical Things} 

\subsection{Opportunities of integrating blockchain with edge intelligence}

The integration of blockchain and edge intelligence can bring mutual benefits to each other. First, blockchain can improve the trustworthiness of edge intelligence. Second, AI algorithms running on edge devices can also enhance blockchain systems. 

Despite advances in AI technologies, there are emerging security and privacy concerns in AI algorithms as well as their-fostered cloud intelligence and edge intelligence. The security threats to AI include stealing, modifying, and even contaminating (or poisoning) the training data~\cite{Miller:Proc.IEEE2020}, while most AI algorithms (especially for deep learning (DL) algorithms) require substantial training on massive data. Moreover, there are also rising concerns on the accidentally or intentionally privacy leaks of IoMT data. The advent of blockchain brings opportunities to address both the security and privacy concerns of AI. In particular, blockchain can offer effective authentication and authorization on the sensory data during data sharing. Any modifications can be detected on time due to the immutability of the blockchain. Moreover, authentication and authorization can be enforced on either edge devices or cloud-service providers through blockchain systems. 

On the other hand, AI can greatly enhance the blockchain ecosystem. Massive data have been generated from diverse blockchain systems. AI-empowered blockchain data analysis can help to extract valuable information and also strengthen blockchain itself. In particular, data analysis of blockchain data can be beneficial to discover the possible defects, predict the failures, and determine the bottleneck factors affecting the system throughput. Consequently, intelligent countermeasures can be made. For example, it is shown in~\cite{PZheng:OJCS2020} that log data analysis can discover the possible faults in the blockchain. Moreover, data analytics on smart contracts (running on the blockchain) can help to detect and identify the vulnerable contract codes, thereby putting forth effective countermeasures. 

\subsection{Architecture of blockchain-empowered edge intelligence for IoMT}

Fig.~\ref{fig:blockchainEI} presents an architecture of blockchain-empowered edge intelligence for IoMT. In particular, the integration of blockchain with AI algorithms can provide the trusted intelligence in three different levels (from bottom to top): 1) device level, 2) edge level, and 3) cloud level. 

\subsubsection{Trusted device intelligence}

\begin{figure*}[h]
\centering
\subfigure[Monitoring and tracing COVID-19 pandemic origin]{
\includegraphics[width=5.8cm]{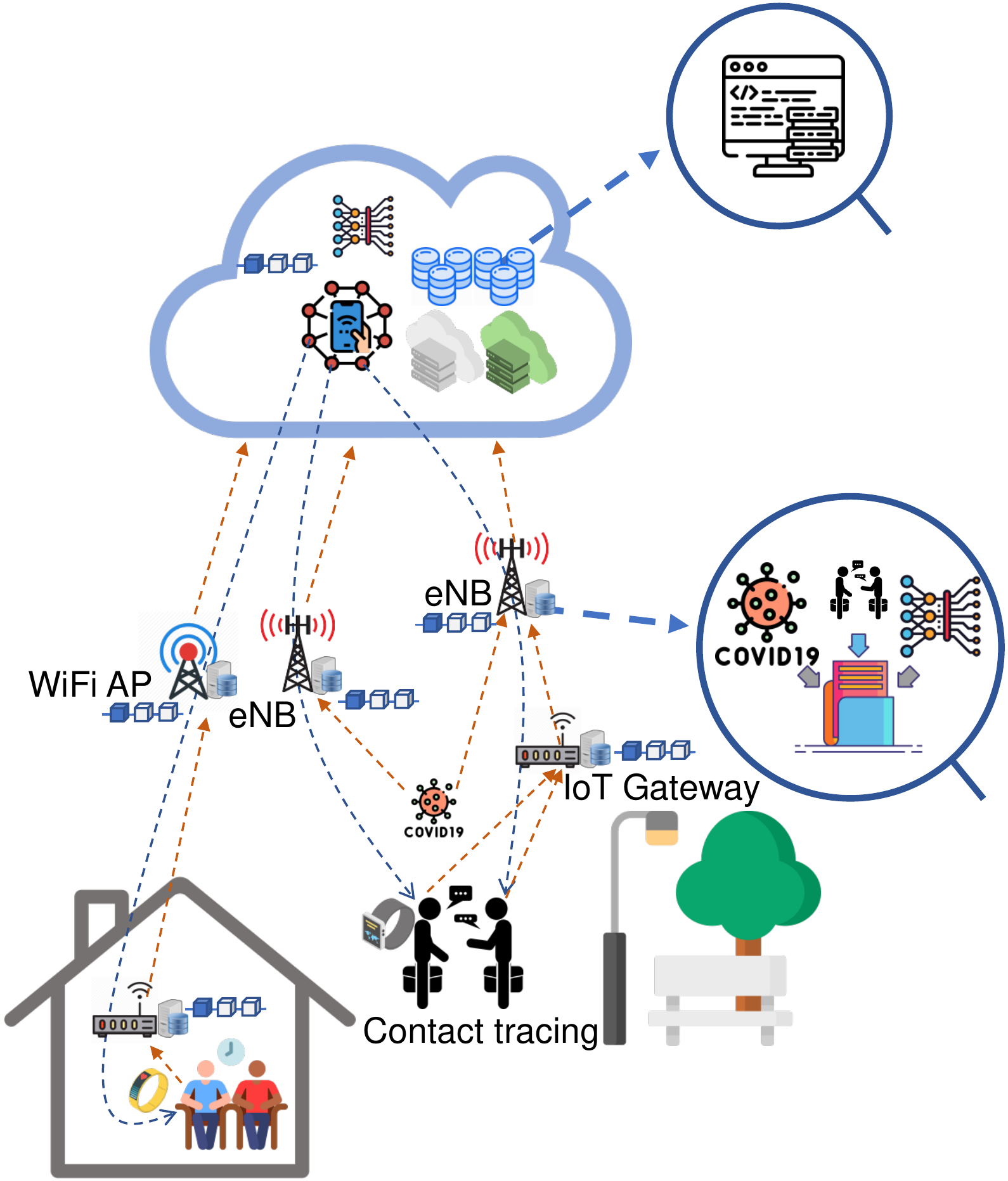}
\label{fig:tracing}
}%
\subfigure[Supply chain of injectable medicines]{
\includegraphics[width=5.8cm]{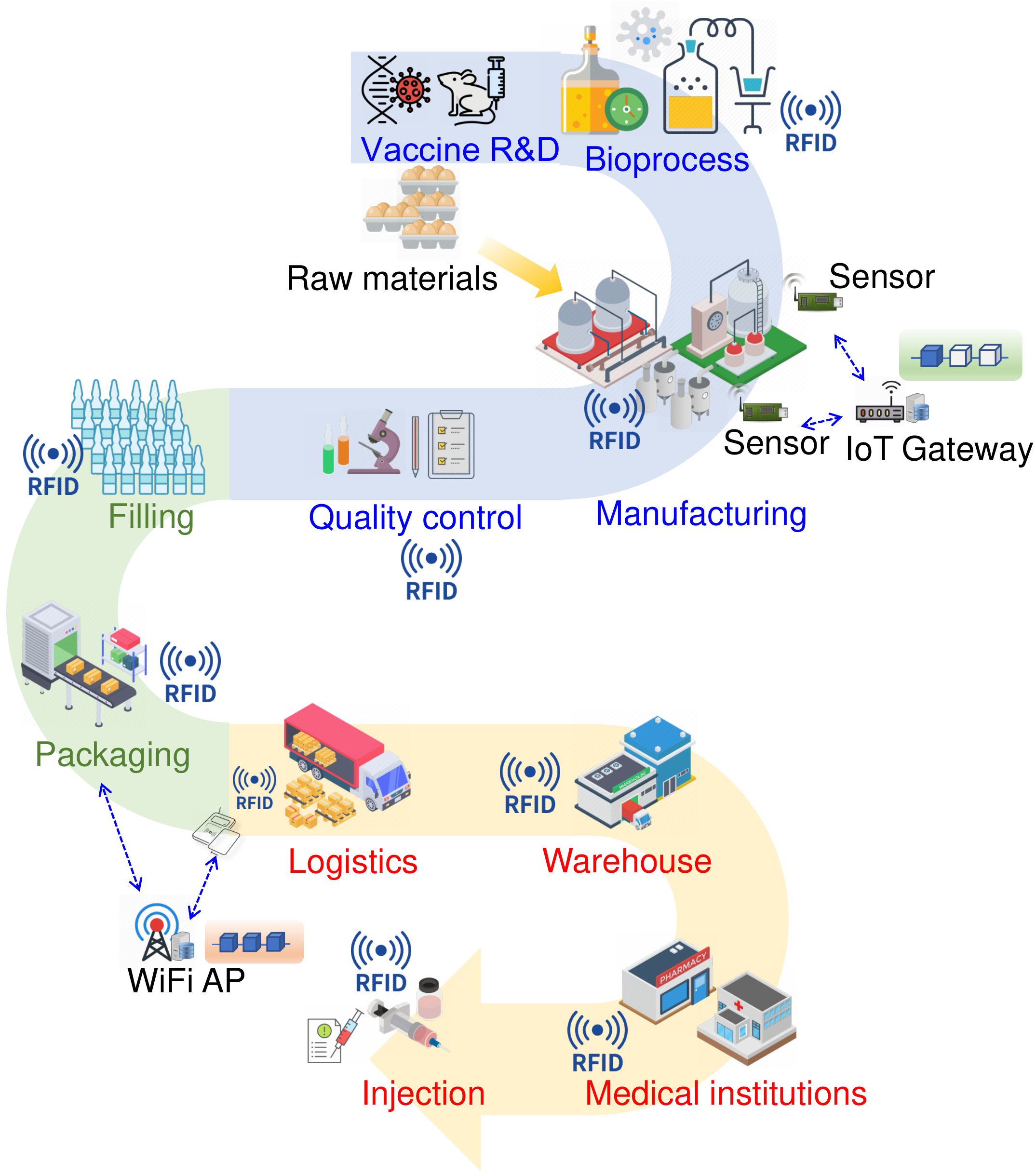}
\label{fig:vaccine-chain}
}%
\subfigure[Telemedicine and remote healthcare services]{
\includegraphics[width=5.8cm]{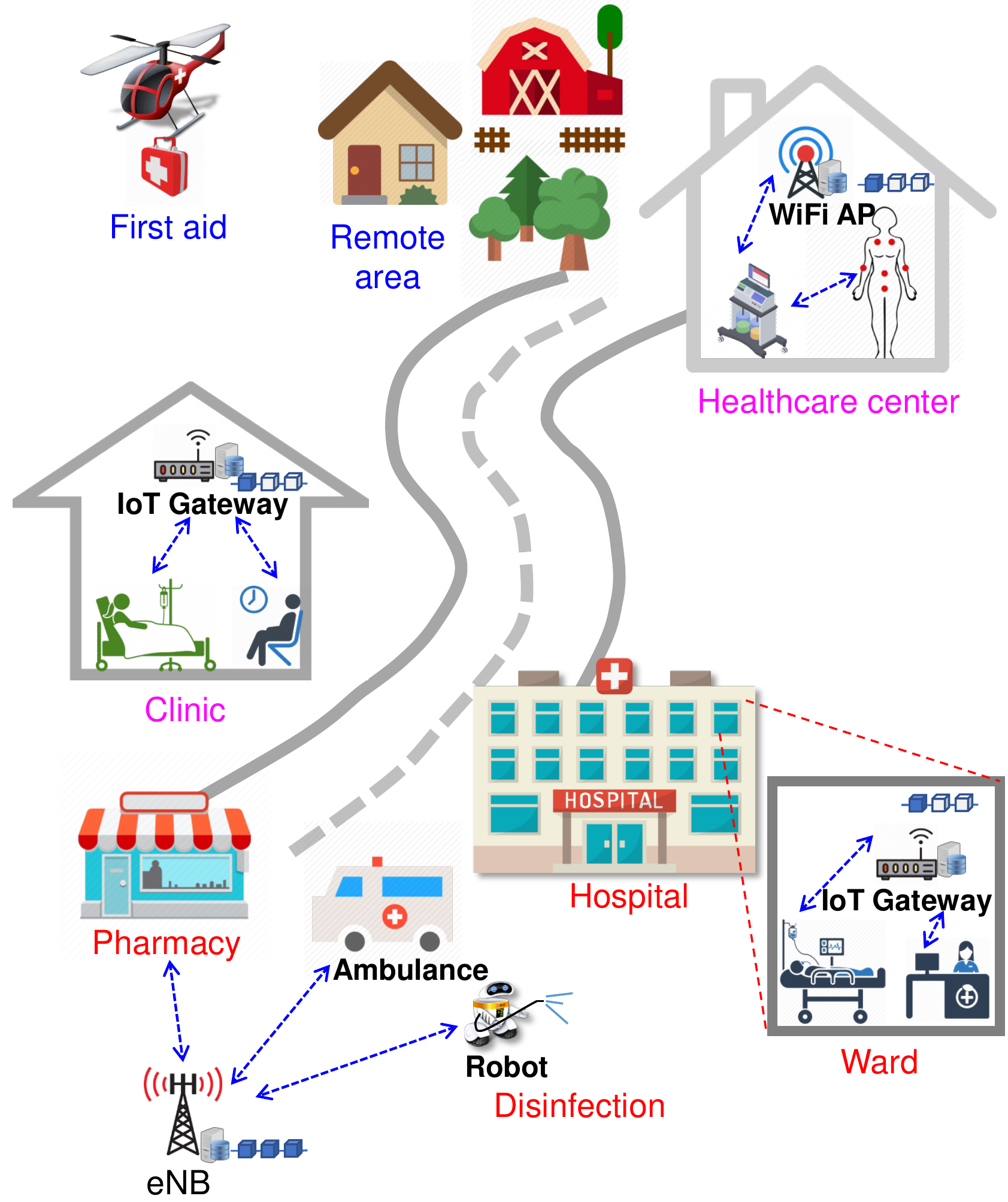}
\label{fig:telemedicine}
}%
\caption{Solutions of Blockchain-empowered Edge Intelligence to IoMT against COVID-19}
\label{fig:solutions}
\end{figure*}

There are diverse IoMT devices including sensors, Radio Frequency Identification (RFID) tags, thermometers, wristbands, and thermal cameras at the device level. Due to the resource limitation, IoMT devices typically cannot store and process massive IoMT data. Moreover, IoMT devices have limited data protection capability, since computationally-complex cryptographic algorithms are not applicable to them. However, edge computing nodes and remote cloud servers can complement IoMT devices in computing and storage capability. For example, IoMT data can be uploaded to nearby edge computing nodes or outsourced to remote cloud servers for further data processing, storage, and analysis. Moreover, the integration of blockchain technologies with IoMT devices can assure the trustworthy management of IoMT devices. In particular, blockchain can facilitate authentication and access control management of IoMT devices. In addition, blockchain can also ensure the anonymity of IoMT devices, since IoT devices only use the generated addresses to interact with each other in the system.

\subsubsection{Trusted edge intelligence}

Edge computing nodes are typically deployed at eNB or gNodeB (gNB in 5G systems), Wi-Fi APs, and IoT gateways, in proximity to IoMT devices, which typically connect with eNB, gNB, APs and IoT gateways via wireless links. 
Compared with IoMT devices that have limited computing capability, edge computing nodes can temporarily store (or cache) and process IoMT data due to the enhanced computing capability and storage capacity~\cite{Abdulsalam:TNSE20}. Meanwhile, lightweight machine learning algorithms have recently been adopted to analyze IoMT data to extract useful information, such as patient symptom identification, contact tracing, and recognizing high-fever infected persons. In contrast to remote cloud computing services, which are far from IoMT devices, edge computing facilities can process time-sensitive tasks. Moreover, edge computing nodes can also protect data privacy since IoMT data can be stored locally at edge computing nodes, at which lightweight cryptographic schemes are also deployed to protect confidential data. Meanwhile, offloading time-sensitive tasks to edge computing nodes can further assure real-time requirements. The integration of edge computing with blockchain can further strengthen the trustworthiness of edge computing nodes in IoMT systems, due to the immutability and the enhanced security brought by blockchain~\cite{dai2020blockchainenabled}.

\subsubsection{Trusted cloud intelligence}

Cloud servers and data centers are mainly responsible for computationally-complex tasks and massive data storage. In particular, DL algorithms can be used to analyze patient medical data so that healthcare practitioners (e.g., medical doctors) can easily identify the symptoms, trace the sources, and make appropriate therapies. However, DL algorithms have stringent demands on the computing facilities, e.g., Tensor Processing Units or Graphics Processing Units to fasten the training process. Meanwhile, DL models also require substantial training on massive IoMT data to achieve outstanding performance (e.g., prediction accuracy). Since IoMT devices and edge devices cannot fulfill such stringent computing requirements, the common practice is to outsource IoMT data to cloud servers, which are often owned by third parties (e.g., MS Azure, Amazon cloud, Alibaba cloud services). Uploading IoMT data to untrusted third parties may intentionally or accidentally leak the data privacy. The introduction of blockchain technologies can protect data privacy and guarantee the trust of cloud intelligence through the effective authentication and authorization of IoMT data. Moreover, the immutability of blockchain can also ensure non-falsifiable IoMT data when it is outsourced to cloud-service providers.

\section{Solutions of Blockchain-empowered Edge Intelligence to IoMT against COVID-19} 

Blockchain-empowered edge intelligence for IoT can solve the challenges posed by COVID-19. Fig.~\ref{fig:solutions} summarizes the solutions. We next elaborate them in details.

\subsection{Monitoring and tracing COVID-19 pandemic origin}
We take monitoring and tracing COVID-19 pandemic origin as a case study to illustrate how the proposed blockchain-empowered edge intelligence works. This case study includes two parallel processes: contact tracing and aerosol surveillance, as shown in Fig.~\ref{fig:tracing}. In the process of contact tracing, a number of trusted IoMTs with diverse wireless connectivity technologies, such as Bluetooth, LoRa, near-field communication (NFC), 4G/LTE/5G, and Wi-Fi are involved in the contact measurement by virtue of localization and positioning techniques. The trusted edge intelligence is in place to improve the accuracy of contact measurement. The signals of close contact IoMTs, e.g., less than 2 meters, will be collected and uploaded to the trusted cloud. The cloud intelligence provides a fast lookup service to quickly find a sequence of close contact IoMTs if a COVID-19 case is confirmed, and the cloud intelligence provides effective data analytics services to identify the origin of this confirmed case through the contact tracing data. All the affected IoMTs will then be notified~\cite{rohmetra2021ai}.

In the process of aerosol surveillance, the related IoT devices (such as Vortex IoT's SalixAir\footnote{\url{https://www.vortexiot.com/products/air-quality-monitoring-aqm/}}) at the device level monitor the contaminated environment and work with trusted edge intelligence to detect aerosol transmission of COVID-19. In the event of a positive detection, the results will be stored in the trusted cloud for further processing, and the IoMTs in the affected region will be alerted. Contact tracing and aerosol surveillance are two intertwined processes to achieve more accurate identification of the origin. The proposed blockchain-empowered edge intelligence provides intelligence at the device level, edge level, and cloud level to carry out data analytics and perform decision making. In addition, the blockchain-empowered capability ensures the transparency of signal data storage and usage as well as data processing with the promise of ensuring personal data anonymization.

\subsection{Traceable supply chain of injectable medicines and COVID-19 vaccines}

Vaccines are expected to be the most effective medicines to prevent the outbreaks of COVID-19. Many countries have devoted huge efforts to vaccine development. There are more than 150 COVID-19 vaccines in development, many of which are in the final test stage across the world. However, the massive production of COVID-19 vaccines may cause a shortage of other critical injectable medicines (e.g., influenza vaccines) considering billions of COVID-19 vaccine demands. Thus, it is a necessity to establish traceable supply chains of injectable medicines and vaccines to maintain enough supply of critical medicines. 

Blockchain-empowered edge intelligence for IoMT can also address this emerging issue. In particular, IoMT deployed in the entire pharmaceutical supply chain can trace the productions of medicines in each phase, as shown in Fig.~\ref{fig:vaccine-chain}. For example, it is reported\footnote{\url{https://www.sandoz.com/news/media-releases/sandoz-launches-its-first-rfid-tagged-critical-injectable-medicines-kit-check}} that Sandoz, an American pharmaceutical company launched a new pharmaceutical supply chain armed with RFID tags so that hospitals can trace the inventory of three injectable medicines from production lines to warehouses. The integration of IoMT with blockchain can further improve the traceability and immutability of medicines in pharmaceutical supply chains. Meanwhile, AI technologies as well as edge intelligence can analyze the massive data of pharmaceutical supply chains so as to optimize the manufacturing process and forecast the coming demands (e.g., a new COVID-19 vaccine).

\subsection{Telemedicine and remote healthcare services}

The outbreaks of COVID-19 brought not only heavy loads to public medical institutions but also revealed the brittleness of medical and healthcare systems. For example, the strong transmission capability of COVID-19 and long incubation period are also the root cause of the COVID-19 pandemic. Meanwhile, the average recovery period of patients infected with COVID-19 is from 6 to 41 days in contrast to the influenza with a recovery period within one week. As a consequence, hospitals have been overcrowded and medical practitioners have been overloaded. Many elderly people and patients with chronic diseases cannot get immediate and proper healthcare services.

Blockchain-empowered edge intelligence also brings solutions to the healthcare crisis caused by COVID-19 pandemic, as shown in Fig.~\ref{fig:telemedicine}. First, the recent technological advances promote the miniaturization of diverse wearable IoMT devices, such as body sensors, sphygmomanometers, glucose meters, and smart bands, which can be connected to remote hospitals or healthcare centers with the provision of real-time monitoring on patients and elderly people. In addition, robots and drones in IoMT systems also play an important role in healthcare digitalization. For example, disinfectant robots or drones can be deployed to disinfect hospitals or contaminated areas so as to reduce the chances of healthcare workers being infected. Moreover, drones can be used to deliver medicines or medical equipment to the remote locked-down area. In both the above cases, IoMT systems need to fulfill both privacy and real-time requirements, which can be addressed by blockchain-empowered edge intelligence. On the one hand, blockchain can effectively protect the privacy of patients' data via enhanced security and pseudonymity of blockchains. On the other hand, edge intelligence can dynamically allocate the resources (e.g., bandwidths and network slices) to fulfill various demands~\cite{XCheng:JSAC20}.

\section{Research Challenges and Open Issues} 
Despite the potential of blockchain-empowered edge intelligence, there are a number of challenges and open issues to be addressed.

\subsection{The role of edge intelligence} Edge intelligence plays a key role in the proposed blockchain-empowered edge intelligence for IoMT. It essentially has three roles. The first is to provide intelligence on enhancing the accuracy of digital contact tracing. The second is to provide intelligence on data analytics and decision making of aerosol surveillance. The third role is to provide intelligence on integrating and fusing the results from the first two roles and make effective identification of the origin and a sequence of affected IoMTs. There are many challenges that need to be tackled in making each role. For example, in the first role, there are many wireless technologies that have been used in contact tracing by virtue of distance measurement. How to provide a unified service at the edge intelligence to improve the accuracy of distance measurement is a challenge. In the second role, how to make an accurate and online detection as well as data analytics in the presence of the changing contaminated environment is hard. In addition, making the third role itself is not an easy task.

\subsection{Offloading} The cooperation among trusted device intelligence, trusted edge intelligence, and trusted cloud intelligence is a necessity in the proposed blockchain-empowered edge intelligence for IoMTs. Offloading is therefore an essential mechanism in this framework. On the one hand, offloading decisions shall consider different metrics at different levels. For example, energy efficiency is an important factor in the device level. On the other hand, necessary local data processing is crucial in reducing the amount of data to be stored in the blockchain, which is an important consideration in terms of sustainability of the proposed framework. For example, light-weight machine learning algorithms can enable device intelligence on deciding the collection of only useful data, e.g., only useful Bluetooth signals are collected.

\subsection{Scalability of blockchain}
Despite immutability, enhanced security, and decentralization of blockchain, many existing blockchain systems are suffering from poor scalability, exhibiting low transactional throughput and high latency (between a transaction being submitted and being confirmed). There are several possible solutions to improve the scalability of blockchains: 1) designing more scalable consensus algorithms, 2) exploring off-chain solutions, 3) resorting to new blockchain structures. With respect to 1), more scalable consensus algorithms include Proof-of-authority (PoA), Practical Byzantine Fault Tolerance (PBFT), Delegated Proof of Stake (DPoS), and their variants. For example, EOSIO with DPoS can reach nearly 1,000 transactions throughput per second (TPS). The main idea of off-chain solutions is to process transactions outside the blockchain, consequently reducing the loads at the main chain. In addition, new blockchain structures like directed acyclic graphs (DAG) were proved to be cost-efficient~\cite{YWu:IoTJ20}.

\subsection{Future AR/VR-based medical applications}

Future telemedicine may support many emerging medical applications based on virtual reality (VR)/augmented reality (AR) technologies. AR/VR-based medical applications include telesurgery, remote surgical planning, training, and education. For example, high-resolution medical images can be reconstructed into 3D medical images via AR and computer vision (CV) algorithms. The ultra-precise 3D medical images can then be used for surgeons to have the better diagnosis, make surgery plans, and offer surgical training to junior surgeons. Integrated with other haptic simulators or tools (e.g., haptic gloves) and AR/VAR technologies, surgeons can mimic surgeries so as to practice and sharpen their surgery skills. These AR/VR-based medical applications have stringent requirements on the reliability, latency, and bandwidth of communication networks. The emerging 5G technologies may fulfill these requirements with the provision of enhanced Mobile BroadBand (eMBB) and ultra-reliable low latency communications (URLLC). However, it is still challenging to fully support telesurgery applications since such complicated medical applications involve different types of 5G services, e.g., eMBB services to transmit ultra-precise 3D images while URLLC to support surgical operations. The convergence of network slicing technology and AI algorithms may potentially solve this problem \cite{XCheng:JSAC20}. In particular, different network slices may be allocated for different medical applications while AI algorithms can help to allocate network resources in a flexible manner.

\section{Conclusion}

This article presented blockchain-empowered edge intelligence for IoMT for combating COVID-19. We first re-visited IoMT technologies as well as the challenges of IoMT. We elaborated blockchain-empowered edge intelligence to address the challenges of IoMT and offer solutions of IoMT to COVID-19. We also outlined potential challenges and open research issues in blockchain-empowered edge intelligence for IoMT in the context of COVID-19.

\bibliographystyle{IEEEtran}
\bibliography{ref_bib}

\begin{IEEEbiographynophoto}{Hong-Ning Dai}
[Senior Member] is currently with Faculty of Information Technology at Macau University of Science and Technology as an associate professor. He obtained the Ph.D. degree in Computer Science and Engineering from Department of Computer Science and Engineering at the Chinese University of Hong Kong. His current research interests include Internet of Things and blockchain technology. He has served as associate editors/editors for IEEE Transactions on Industrial Informatics, IEEE Systems Journal, Connection Science, Ad Hoc Networks, and IEEE Access. He is a senior member of Association for Computing Machinery (ACM). 
\end{IEEEbiographynophoto}

\begin{IEEEbiographynophoto}{Yulei Wu}
[Senior Member] is a Senior Lecturer with the Department of Computer Science, College of Engineering, Mathematics and Physical Sciences, University of Exeter, United Kingdom. He received the B.Sc. degree (First Class Honours) in Computer Science and the Ph.D. degree in Computing and Mathematics from the University of Bradford, United Kingdom, in 2006 and 2010, respectively. His expertise is on intelligent networking, and his main research interests include computer networks, networked systems, and network security and privacy. He is an Editor of IEEE Transactions on Network and Service Management, IEEE Transactions on Network Science and Engineering, Computer Networks (Elsevier) and IEEE Access. He is a Fellow of the HEA (Higher Education Academy).
\end{IEEEbiographynophoto}


\begin{IEEEbiographynophoto}{Hao Wang}
[Senior Member] is an Associate Professor in the Department of Computer Science in Norwegian University of Science and Technology, Norway. He has a Ph.D. degree and a B.Eng. degree, both in computer science and engineering, from South China University of Technology. His research interests include big data analytics, industrial internet of things, high performance computing, and safety-critical systems. He served as a TPC co-chair for IEEE DataCom 2015, IEEE CIT 2017, ES 2017, and IEEE CPSCom 2020, a senior TPC member for CIKM 2019, and reviewers/TPC members for many journals and conferences. He is the Chair for Sub TC on Healthcare in IEEE IES Technical Committee on Industrial Informatics. He is also a member of the ACM.
\end{IEEEbiographynophoto}

\begin{IEEEbiographynophoto}{Muhammad Imran} 
[Senior Member] is working as an associate professor in the College of Applied Computer Science, King Saud University. His research interests include mobile and wireless networks, Internet of Things, cloud and edge computing, and information security. He has published more than 200 research articles in reputable international conferences and journals. His research is supported by several grants. He serves as an associate editor for many top ranked international journals. He has received various awards.
\end{IEEEbiographynophoto}

\begin{IEEEbiographynophoto}{Noman Haider} is a working as a lecturer in the College of Engineering and Science at Victoria University, Australia. He received his Ph.D. in engineering from the University of Technology Sydney, Australia in 2019. His research interests include mobile and wireless networks, Internet of Things, data science, and information security.
\end{IEEEbiographynophoto}




\end{document}